# The Sensemaking-Coevolution-Implementation Theory of Software Design

Paul Ralph

Lancaster University

paul@paulralph.name

**Abstract**

Understanding software design practice is critical to understanding modern information systems development. New developments in empirical software engineering, information systems design science and the interdisciplinary design literature combined with recent advances in process theory and testability have created a situation ripe for innovation. Consequently, this paper utilizes these breakthroughs to formulate a process theory of software design practice: Sensemaking-Coevolution-Implementation Theory explains how complex software systems are created by collocated software development teams in organizations. It posits that an independent agent (design team) creates a software system by alternating between three activities: organizing their perceptions about the context, mutually refining their understandings of the context and design space, and manifesting their understanding of the design space in a technological artifact. This theory development paper defines and illustrates Sensemaking-Coevolution-Implementation Theory, grounds its concepts and relationships in existing literature, conceptually evaluates the theory and situates it in the broader context of information systems development.

# The Sensemaking-Coevolution-Implementation Theory of Software Design

## 1 INTRODUCTION

Cross (1992) pointed out that "At the moment, we seem to have a fairly rich picture of design thinking, but we lack a successful, simplifying paradigm of design thinking … The lack of an adequate, simplifying paradigm is perhaps something which inhibits the transfer of knowledge from research in practice and education" (p. 9). To address this gap, this paper proposes a software design process theory, which attempts to incorporate many key findings from empirical research on design, especially software and information systems design.

The importance of this step is entwined with the emerging bifurcation of design literature. Many authors (e.g., Dorst and Dijkhuis 1995; Franz and Robey 1984a; Ralph 2011) argue that design research is divided into two paradigms. The dominant paradigm, the "Rational Model of Design", primarily consists of goals, desiderata, constraints, resources, a utility function and a tree of design decisions that the designer searches for an optimal or satisfactory candidate (Brooks 2010). Although Brooks does not comprehensively define the Rational Model, he indicates three formulations thereof:

1. the mechanical-engineering view of design as a systematic, orderly process, as described by Pahl and Beitz (1996);

2. the artificial-intelligence view of design as a search for satisfactory alternatives given goals and constraints, by a designer exhibiting "procedural rationality", as formulated by Simon (1996);

3. the managerial view of design as a sequence of weakly-coupled phases (the Waterfall Model); however, more the straw-man forward-only model Royce (1970) was attacking than the more sophisticated iterative model that he was proposing.

These formulations sharply contrast with the subordinate "Empirical Model of Design" (Ralph 2011). This model primarily consists of a fallible, emotional designer who, faced with ambiguous, conflicting goals and objectives, explores a solution space using design concepts in an improvisational, iterative manner. Formulations of this view include:

1. the view of designer as a "reflective practitioner" alternating between problem framing, adjusting a design concept and evaluating the adjustment's consequences (Schön 1983);

2. the view of the designer as a creative agent whose attention oscillates between an ill-defined problem concept and a tentative solution concept, gradually developing an understanding of both (Cross et al. 1992; Dorst and Cross 2001);

3. "segmented institutionalism", a theoretical perspective that assumes intergroup conflict on goals and the supremacy of emotional considerations over efficiency (Kling 1980).

The Rational Model powerfully frames the academic discourse around software design, information systems design and engineering. It comprises the assumptions underlying research (e.g., Ewusi-Mensah 2003; Fitzgerald 2006), methods (e.g., Jacobson et al. 1999), textbooks (e.g., Baltzan and Phillips 2008; Kroenke et al. 2010; Laudon et al. 2009), curriculum guidelines (e.g., Joint Task Force on Computing Curricula 2004; Topi et al. 2010), and standards (e.g., IEEE 1998). Brooks argued that "the need to communicate and the nature of academic instruction means that there will be a dominant model of the design process" (2010, p. 52). However, the dialectic process of comparing, contrasting, integrating or moving beyond the Rational / Empirical distinction is impeded by the absence of a simplifying process theory on the Empirical side, as lamented by Cross (above). Put another way, long-recognized problems with the Rational Model (cf. Brooks 2010;

Curtis et al. 1992; Gladden 1982; McCracken and Jackson 1982) have remained for decades partly because no comprehensive alternative with which to challenge it was available. Therefore, formulating a comprehensive process theory of software design practice consistent with the Empirical Model may stimulate innovation, lead to more broadly-applicable theory and further our collective understanding of practice in information systems development (ISD) and software engineering (SE). Consequently, the purpose of this paper is as follows.

> ***Purpose:*** *to formulate a process theory of software design practice, which explains how complex software systems are created by collocated software development teams in organizations, consistent with the Empirical Model of Design.*

Here, process theory refers to an explanation of how and why an entity changes and develops (Van de Ven 2007; Van de Ven and Poole 1995) rather than a causal chain (Markus and Robey 1988; Sabherwal and Robey 1995). A software design process theory, then, is an explanation of how and why a software system changes and develops. The term, theory of *practice*, is used to denote concern with how software is developed in reality, including both effective and ineffective practices, rather than positing an idealized process or prescribing good ways of designing.

Such theories are important in IS in several ways. Benbasat and Zmud (2003) argued that the design of IT artifacts is central to the IS discipline. Furthermore, recent years have witnessed a surge of interest in design science, which includes two research streams (Hevner and Chatterjee 2010a) – a research method where knowledge is gained by building innovative artifacts (Hevner et al. 2004) and the study of the nature of design work (Löwgren and Stolterman 2004). This paper clearly fits with the second stream, more specifically within software design science, the philosophical, theoretical and empirical study of software creation and modification including its phenomenology, methodology and causality.

Clearly, software systems are not equivalent to information systems – some IS have no software (e.g., filing cabinets) and some software systems are not IS (e.g., Pac-Man). Practically speaking, however, the centrality of software to modern IS, including enterprise systems, makes the design of each intermingled. Therefore, understanding software development is crucial to understanding IS development, which in turn is central to both the IS discipline and the second stream of design science research.

Furthermore, for the purposes of this paper, the phenomenon of interest may be defined as the design of *complex* software systems. Here a *complex system* is specifically one exhibiting emergent (unpredictable) behavior (Waldrop 1992) – complex systems are not necessarily large or composed of many components. The level of analysis varies from individual to team. Meanwhile design refers to an agent creating a specification of an object, which is intended to accomplish goals in a particular environment using a set of primitive components and subject to constraints (Ralph and Wand 2009). Moreover, following Freeman and Hart (2004), "here, design encompasses all the activities involved in conceptualizing, framing, implementing, commissioning, and ultimately modifying complex systems—not just the activity following requirements specification and before programming, as it might be translated from a stylized software engineering process" (p. 20).

Developing a software design process theory is particularly appropriate at this time for three reasons. First, increased interest in design (Hevner and Chatterjee 2010b) and process theory (Gregor 2006) in IS has whet that field's appetite for design process theory; meanwhile numerous calls for more theorizing in SE (Ralph et al. 2013, Johnson et al. 2012) demonstrate that field's appetite for design process theory. Second, Brooks' (2010) elucidation of the Rational Model's limitations provides a clearer problem contextualization than had been previously available. Third, the ongoing crisis of high software project failure rates and overruns (Ewusi-Mensah 2003; Molokken and Jorgensen 2003; Standish Group 2009) points to the ubiquity and severity of this problem in practice. In summary, the higher standard of knowledge reached by recent developments in MIS and elsewhere permits a greater leap in both generalizability and parsimony than was previously plausible. This will become more clear as the proposed theory unfolds below.

Next, I review existing process theories from design literature (§2). Section Three introduces, justifies and evaluates the proposed process theory, Sensemaking-Coevolution-Implementation Theory. I then discuss the proposed theory's relationships to ISD literature (§4). Section Five summarizes the paper's contributions and concludes with implications for research, practice and education.

# 2 EXISTING DESIGN PROCESS THEORIES

## 2.1 Review Structure and Sampling

As design literature is spread across many fields, journals, books and other outlets, I conducted a search-based review rather than a comprehensive analysis of a particular set of publications. The review focused on three overlapping fields – information systems (e.g., MIS Quarterly), software engineering (e.g., the International Conference on Software Engineering) and the interdisciplinary design literature (e.g., Design Studies). As the review spread through bidirectional citation search, examining both what a work references and what references it, some particularly influential work from architecture, engineering and sociology was drawn in.

ISD literature is predominately concerned with *methods* for developing systems and managing the development process (Truex et al. 2000) and methods literature is predominately normative / prescriptive (Wynekoop and Russo 1997). Methods are also called methodologies and processes with different authors denoting slightly different phenomena with these terms. For the purposes of this paper, a *method* is a collection of prescriptions for developing software / information systems (Software Development Methods; SDMs) or managing development projects (Project Management Frameworks; PMFs). Types of prescriptions include practices, techniques, tools and phase models. Methods are distinct from both process models and process theories. A process model is "an abstract description of an actual or proposed process" (Curtis et al. 1992, p. 76). A *process theory* is an explanation of how and why an entity changes and develops (Van de Ven 2007). A *design process theory* is an an explanation of how and why a *design artifact* changes and develops. A description qualifies as a design process theory if it meets the following three criteria.

1. It posits a formative relationship between a higher level phenomena (design) and several lower-level phenomena (constructing a model, identifying expected behaviors).
2. It includes a causal motor (dialectic, evolutionary, lifecycle, teleological) (Table 1).
3. It includes a claim to universality within a domain.

Table 1. Ideal Types of Process Theories (adapted from Ralph 2010b; Van de Ven and Poole 1995)

| Ideal Type | Proponents | Capsule Description | Event Progression | Contemporary Example |
|---|---|---|---|---|
| Dialectic | (Plato; Hegel; Van de Ven et al. 1995) | Changes result from shifts in power among conflicting entities | Recurrent, discontinuous sequence of conflict and resolution | Behavioral Negotiation Theory (Neal and Northcraft 1991) |
| Evolutionary | (Darwin; Van de Ven et al. 1995) | A population of entities changes as less fit entities expire and remaining entities change and recombine | Recurrent, cumulative and probabilistic sequence of variation, selection and retention | Change in populations of organizations (Carroll and Hannan 1989) |
| Lifecycle | (Markus and Robey 1988; Van de Ven and Poole 1995) | An entity progresses through a series of stages in a predefined sequence | Linear & irreversible sequence of prescribed stages | The Organizational Lifecycle (Kimberly and Miles 1980) |
| Teleological | (Churchman 1971; Singer 1959; Van de Ven and Poole 1995) | An agent purposefully selects and takes actions to achieve a goal | Recurrent, agent-determined sequence of goal setting and action taking | Organizational decision making (March and Simon 1958) |

In summary, methods *prescribe*, process models *describe* and process theories *explain*. More to the point, methods in general are inappropriate foundations for design theories (Vermaas and Dorst 2007) as a method purports to *prescribe an effective way* of creating software while a software design process theory purports to *explain* software creation *in general*, regardless of effectiveness (cf. Gregor 2006). Therefore, a concept may be a method or a theory but not both.

As this paper fundamentally concerns process theory, the following review is limited to process theories, and does not attempt to review methods or process models. One exception is the Waterfall Model, which is included as it is often treated as a process theory.

While I uncovered nothing explicitly labeled "design process theory", I found several "models" and "frameworks" that meet the above criteria. Consistent with the Rational Model are the Waterfall Model (treated as a lifecycle theory), the Basic Design Cycle (lifecycle), the Problem-Design Exploration Model (evolutionary) and the Function-Behavior-Structure Framework (teleological). Only the Selfconscious Process, a teleological theory, was consistent with the Empirical Model. These theories are each discussed below, followed by a exploration of the need for new theory.

## 2.2 The Waterfall Model

The "Waterfall Model" refers to a collection of SDMs that divide software or systems development into a series of phases including *analysis*, *design*, *coding* and *testing* or synonyms thereof. Different versions prescribe different numbers of phases and different degrees of interaction between them. Figure 1 shows three versions: 1) solid arrows show the forward-only version; 2) solid and dashed lines show a version allowing backtracking; 3) solid, dashed and dotted lines indicate a *clique* version, where transition between any two activities is allowed. The original, forward-only version from which the method gets its name was proposed by Royce (1970) as a straw man, which he dismissed as too risky, in favor of a more iterative method.

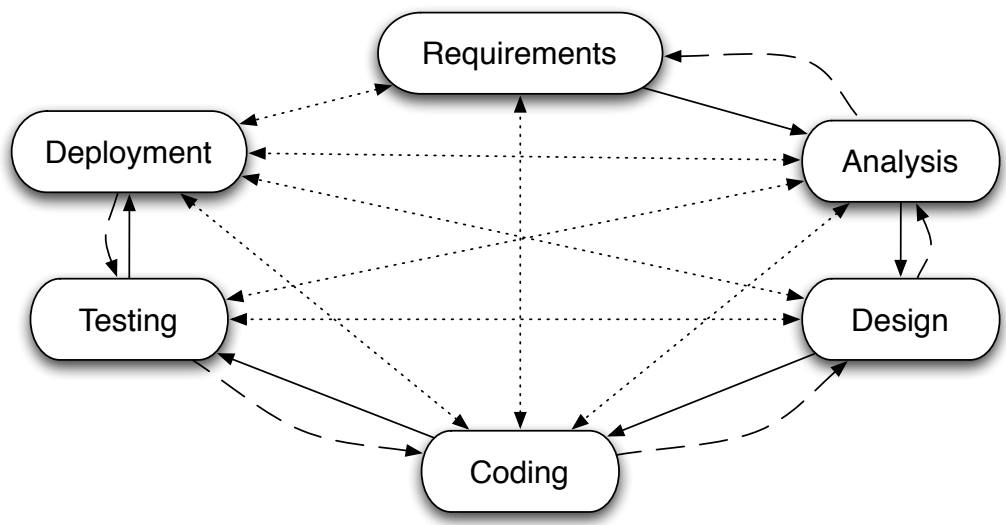

Fig. 1. Generalized Waterfall Model (adapted from Royce 1970)

The view of the Waterfall Model as a method, with or without backtracking, remains common in research literature (e.g., Lee and Xia 2010; Sircar et al. 2001) and textbooks (e.g., Baltzan and Phillips 2008; Kroenke et al. 2010). However, the *clique* version is often treated as process theory in contemporary IS discourse. For example, Fitzgerald (2006) stated that "in conventional software development, the development lifecycle in its most generic form comprises four broad phases: planning, analysis, design, and implementation" (p. 589) and Ewusi-Mensah claimed "regardless of the particular process model an organization may use … every software project will feature: (1) the requirements-definition and functional-specification phase; (2) the design phase; … (3) the implementation or the coding and testing phase; … and (4) the installation, operation, and maintenance phase" (2003, p. 51). In a popular introductory textbook, Laudon et al. (2009) stated that "systems development … consist[s] of systems analysis, systems design, programming, testing, conversion and production and maintenance … which usually take place in sequential order". Other popular textbooks (e.g., Baltzan and Phillips 2008; Kroenke et al. 2010) continue to present systems development based on Waterfall phases. These phases are also explicitly adopted in the official IEEE Guide to the Software Engineering Body of Knowledge (Bourque and Dupuis 2004). Finally, as of May 14, 2012, the Systems Development Lifecycle (SDLC) Wikipedia article states that "A SDLC adheres to important phases that are *essential* for developers, such as planning, analysis, design, and implementation" (italics added). When a text is organized using Waterfall phases or treats them as 'essential' or 'generic', the Waterfall Model is being treated as process theory rather than a SDM.

The Waterfall Model's elevation to de facto process theory (or possibly dogma) is central to Brooks' (2010) assertion that the Rational Model, as expressed by the Waterfall Model, is the dominant view of design. In terms of ideal type, the Waterfall Model most closely resembles a lifecycle process theory – an explanation of how and why an entity changes and develops according to a prefigured, unitary sequence of phases (Van de Ven and Poole 1995).

The Waterfall Model, especially forward-only variations, is attractive in several ways. As it has existed for many years, the phases are well-known. It is easy to understand and provides clear milestones – 'the analysis phase is complete, now onto design'. The structure is consistent with basic methods of problem solving (Polya 1957) – understand the problem, devise a plan to solve it, then execute the plan. It facilitates "a fixed price contract for a specific delivery date" (Brooks 2010, p. 44). More generally, it allows managers to drive development through schedule and budget, regardless of their practicality. It also underlies the division of labor between specialists in analysis, design, testing, etc. as seen in the Unified Process (Jacobson et al. 1999).

However, the Waterfall Model has been extensively criticized as harmful (Brooks 2010; McCracken and Jackson 1982) and incommensurate with observed development practice (Bansler and Bødker 1993; Checkland 1999; Cross et al. 1992; Hartmann 2006; Ralph 2010a; Standish Group 2006; Truex et al. 2000; Zheng et al. 2011).

In summary, the Waterfall Model is a class of SDMs that is often treated as a lifecycle process theory. It remains influential in IS discourse, possibly due to its long history, intuitive appeal and compatibility with management methods, despite some evidence that it is misleading.

## 2.3 The Basic Design Cycle

The Basic Design Cycle (Roozenburg and Eekels 1995) models design as a primarily linear sequence of problem analysis, solution synthesis and simulative testing, culminating in a decision to proceed with the proposed solution "or try again and generate a better design proposal" (p. 92) (Figure 2). However, it meets all three process theory criteria: 1) it relates design to lower-level activities (synthesis, simulation, evaluation); 2) it has a lifecycle causal motor; 3) it includes a claim to universality – Roozenburg and Eekels state that "someone who claims to have solved a design problem has gone through this cycle at least once" (p. 89). Its structure is not unlike the Waterfall Model (above). Like the Waterfall Model, the Basic Design Cycle is consistent with the Rational Model, as evidenced by its separation of analysis from design, assumption of a known, stable problem and general systematicity.

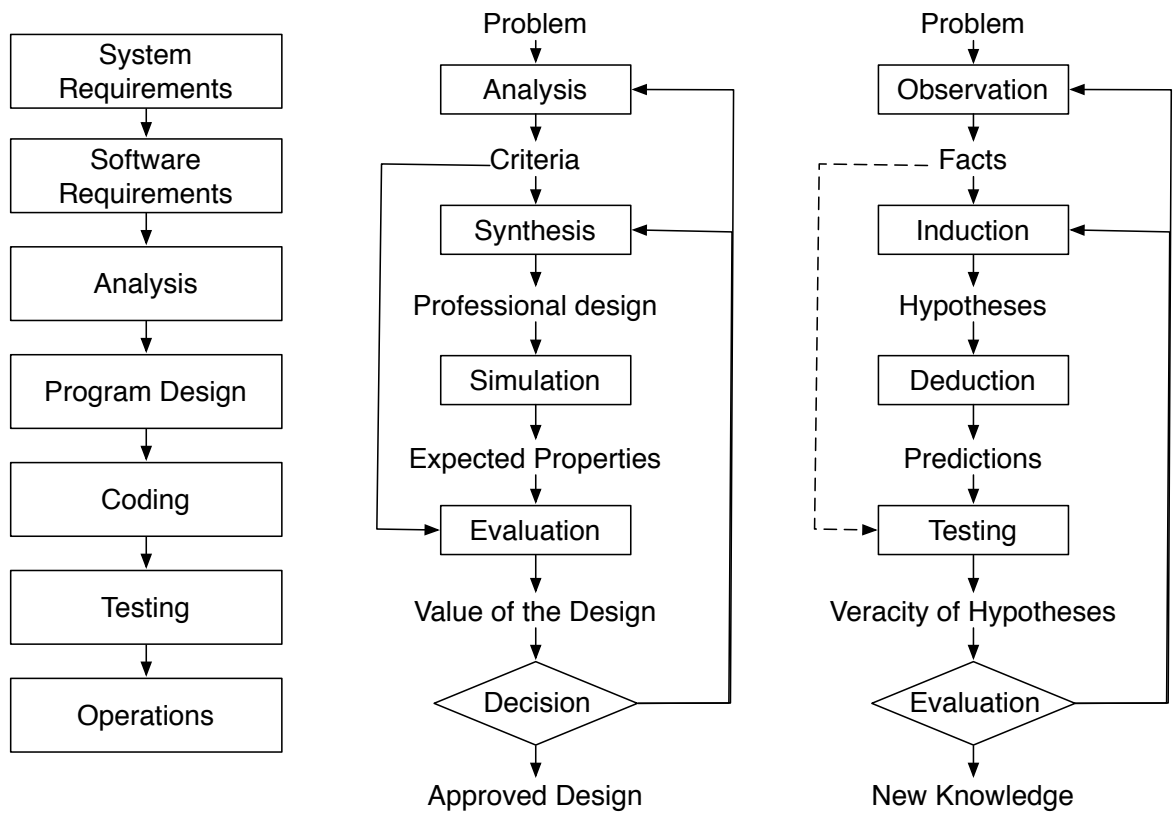

**Fig. 2.** The Basic Design Cycle (centre) with the Waterfall Model (left) and the Basic Cycle of Scientific Inquiry (right), (adapted from Roozenburg and Eekels 1995)

The Basic Design Cycle mostly shares the Waterfall Model's benefits and limitations – it is easy to understand and communicate and is intuitively appealing. However, it appears inconsistent with software development practice (Bansler and Bødker 1993; Checkland 1999; Cross et al. 1992; Hartmann 2006; Ralph 2010a; Standish Group 2006; Truex et al. 2000; Zheng et al. 2011). To be clear, however, as it was devised to explain *engineering* design it can hardly be faulted for not accurately explaining *software* design.

## 2.4 The Problem-Design Exploration Model

The Problem-Design Exploration Model (Maher et al. 1995) embodies a substantially different standpoint, modeling design as two interacting evolutionary systems – problem space *P* and solution space *S*. Distinguishing between problem and design spaces is evident in analytical (e.g., Dorst and Cross 2001), empirical (e.g., Gero and McNeill 1998) and prescriptive (e.g., Checkland 1999) design research. Purao et al. (2002) described "the problem space … as the metaphoric space that contains mental representations of the developer's interpretation of the user requirements" and "the design space" as " the metaphoric space that contains mental representations of the developer's specific solutions" (p. 251-252). From this distinction, Maher et al. (1995) suggest a formal model of exploration intended to describe how an artifact (not necessarily software) may be designed by genetic algorithms (Figure 3).

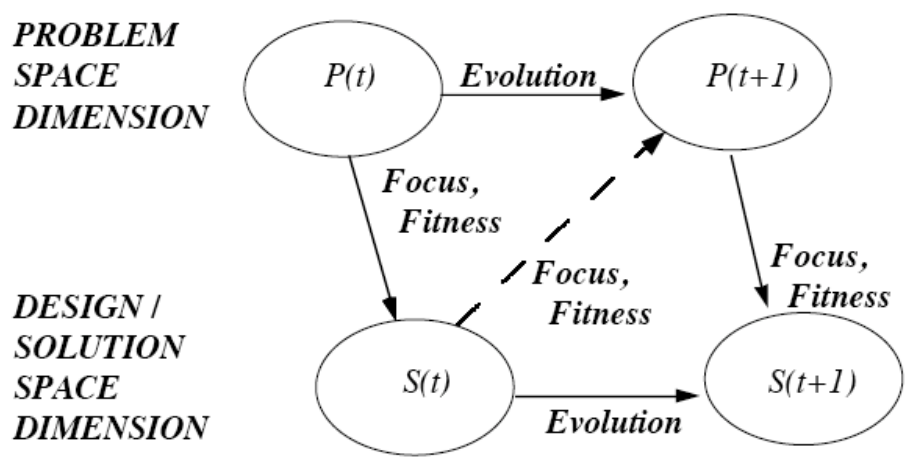

**Fig. 3.** Problem-Design Exploration Model (adapted from Maher et al. 1995)

*Note: P(t) = problem at time t; S(t) = situation at a time t; dashed line indicates situation refocusing problem; diagonal downward movement indicates a search process.*

This constitutes a process theory – it relates design to lower-level activities (evolution, focus, fitness); it has an evolutionary causal motor; Maher et al. position it as a universal explanation within the design-by-genetic-algorithms domain. Moreover, later work applied Maher et al.'s model directly to design by human agents, arguing that coevolution of problem and solution spaces is central to human design processes (Dorst and Cross 2001, p. 434).

Given its explicit focus on evolution, the Problem-Design Exploration Model is closest to the evolutionary ideal type. Furthermore, as it assumes a bounded problem space, operationalized as the system's requirements, and is built on a search metaphor, Maher et al.'s theory is more consistent with the Rational Model.

Important benefits of the Problem-Design Exploration Model include its overall simplicity and its clear elucidation of the coevolution phenomenon observed in field studies of designers (cf. Cross et al. 1992; Schön 1983). However, as the theory was devised to show how genetic algorithms apply to systems design, it is not clear how it translates to human design practice. Specifically, for genetic algorithms, the problem and solution spaces must be represented as a finite number of known dimensions, where for human designers, problem and solution spaces are merely metaphors and often lack such precise definition.

## 2.5 Alexander's Design Process Models

Differentiating *form* (the object being designed) from *context* (the object's environment), prolific architect Christopher Alexander proposed three "possible kinds of design process" (1964, p. 75) (Figure 4). In the "Unselfconscious Process", the designer directly manipulates the design object and other items in the external world to eliminate misfits between form and context (e.g., an igloo dweller directly manipulates the igloo's structure to respond to temperature changes, creating vents when the temperature rises and eliminating them when the temperature falls). In the "Selfconscious Process", the designer works by iterating between "the conceptual picture of the context … and ideas and diagrams and drawings which stand for forms" (p. 75) (e.g., a fashion designer sketches handbags from different angles, changing features based on her ideas about aesthetics). In the third process, which Alexander did not name but may be called the "Formal Process", the designer creates a formal model of the mental pictures using set theory and solves problems using a divide and conquer strategy.

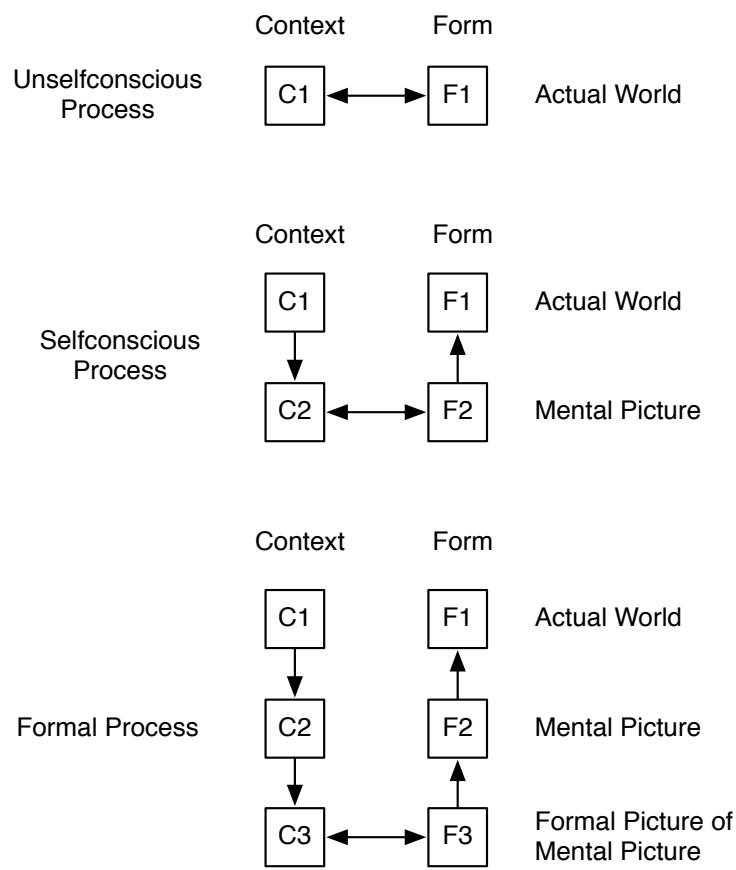

**Fig. 4.** The Selfconscious Design Process (adapted from Alexander 1964)

*Note: arrows indicate interactions, not sequence; Alexander does not specify the exact nature of these interactions.*

These three models are not all process theories. Of the first model, Alexander makes no claim to universality and as it has only one interaction it does not explain design in terms of lower level processes. Of the third model, as Alexander explicitly *prescribed* it to overcome the limitations of the Selfconscious Process, it is a method.

However, the Selfconscious Process does explain design in terms of three lower-level interactions, it has a (teleological) causal motor and Alexander presented it as a model of the design process generally used by architects and other designers. Moreover, it appears consistent with software design – from left to right, the first interaction is analogous to problem analysis, the second to the coevolution of problem and solution spaces discussed above, the third to construction or implementation.

Viewed as a process theory, Alexander's Selfconscious Process would fit the teleological ideal type where an autonomous agent pursues a goal by taking actions in a self-determined sequence and monitors progress (Van de Ven and Poole 1995). Furthermore, it is broadly consistent with the Empirical Model, given its direct interaction with the context and lack of emphasis on planning.

Benefits of the Selfconscious Process include its parsimony and inclusion of problem framing. However, its concepts and relationships need further clarification – Alexander pointed out that the relationship between and, as with the Problem-Design Exploration Model, its was not intended .

## 2.6 The Function-Behavior-Structure Framework

The Function-Behavior-Structure Framework (FBS) (Gero 1990) claims that "the purpose of designing is to transform *function*, F (where F is a set), into a *design description*, D, in such a way that the artifact being described is capable of producing those functions" (Gero 1990, p. 2, original italics). The original FBS (Figure 5) had five artifacts (Table 2) and eight processes (Table 3).

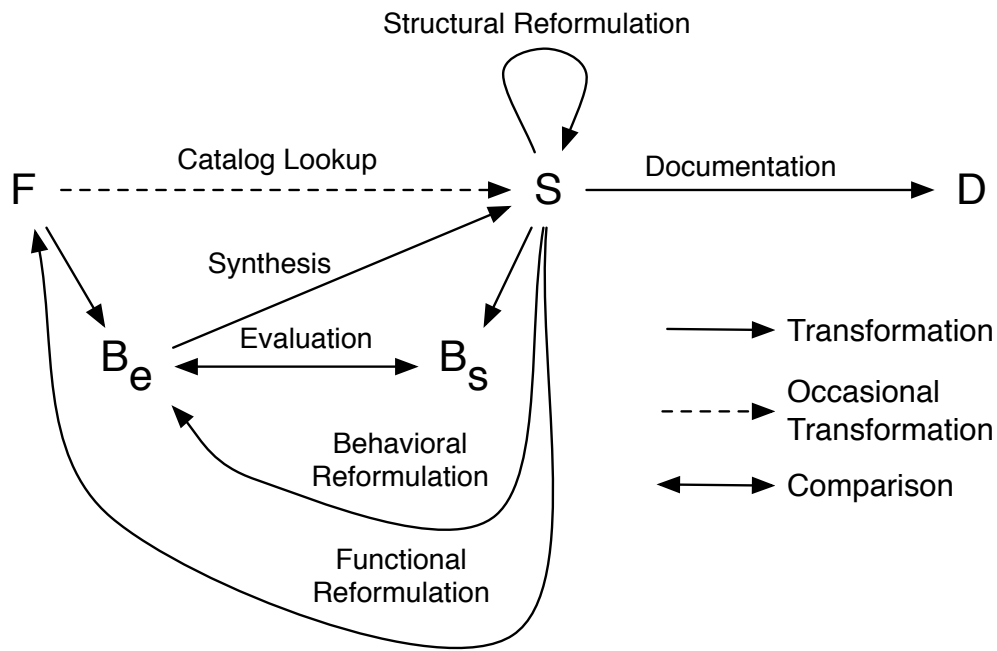

**Fig. 5.** The Function-Behavior-Structure Framework (adapted from Kruchten 2005)

| | Table 2. Artifacts of FBS (adapted from Gero 1990) |
|---|---|
| **Symbol** | **Meaning** |
| Be | expected (desired) behavior of the structure |
| Bs | "the predicted behavior of the structure" (p. 3) |
| D | a graphically, numerically and/or textually represented model that transfers "sufficient information about the designed artifact so that it can be manufactured, fabricated or constructed" (p. 2) |
| F | "the expectations of the purposes of the resulting artifact" (p. 2) |
| S | "the artifact's elements and their relationships" (p. 2) |

Table 3. (Situated) FBS Processes (after Gero and Kannengiesser 2007a)

| Process | Meaning | Situated Activities |
|---|---|---|
| Formulation | deriving expected (desired) behaviors from the set of functions | 1 through 10 |
| Synthesis | "expected behavior is used in the selection and combination of structure based on a knowledge of the behaviors produced by that structure" (p. 3) | 11, 12 |
| Analysis | the process of deriving the behavior of a structure | 13, 14 |
| Evaluation | comparing predicted behavior to expected behavior and determining whether the structure is capable of producing the functions | 15 |
| Documentation | transforming the structure into a design description that is suitable for manufacturing | 12, 17, 18 |
| Structural Reformulation | modifying the structure based on the structure and its predicted behaviors | 9 (possibly driven by 3, 6 or 13) |
| Behavioral Reformulation | modifying the expected behaviors based on the structure and its predicted behaviors | 8 (possibly driven by 2, 5, 14 or 19) |
| Functional Reformulation | modifying the set of functions based on the structure and its predicted behaviors | 7 (possibly driven by 1, 4, 16 or 20) |

Gero and Kannengiesser (2004) updated FBS to include the idea of *situatedness* ("the agent's view of a world changes depending on what the agent does" p. 90), leading to distinction between three "worlds".

> *The external world is the world that is composed of representations outside the designer or design agent. The interpreted world is the world that is built up inside the designer or design agent in terms of sensory experiences, percepts and concepts…The expected world is the world imagined actions will produce (p. 93).*

Gero and Kannengiesser (2004) argued that function, behavior and structure have representations in each world and mapped the original eight FBS processes onto 20 activities across the three worlds. The three-world metaphor, twelve representations, eight processes and 20 activities comprise *Situated FBS* (Figure 6, Table 3).

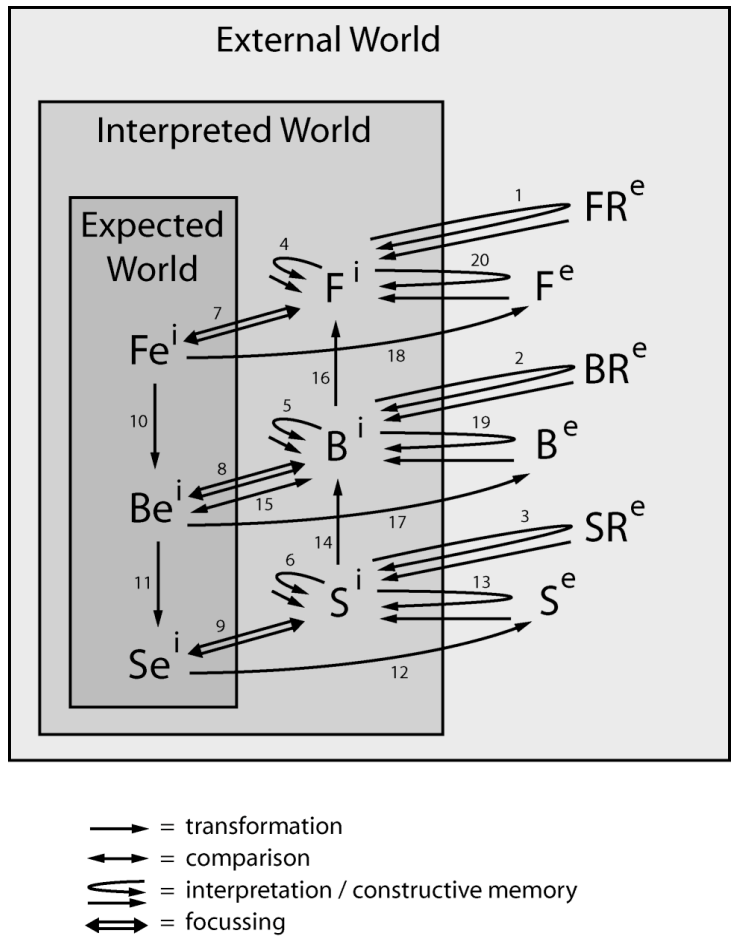

**Fig. 6.** The Situated FBS Framework (after Gero and Kannengiesser 2007a)

*Note: numbers are labels and do not indicate sequence.*

As Situated FBS can be somewhat overwhelming, an informal exposition may help. $FR^e$, $BR^e$ and $SR^e$ are the functional, behavioral and structural requirements provided to the designer. Based on these requirements, the designer produces a structure, $S^i$. $Be^i$ and $Fe^i$ are the behaviors and functions that the designer desires $S^i$ to produce, while $B^i$ and $F^i$ are the behaviors and functions the designer predicts $S^i$ will actually produce. $S^i$ is evaluated by comparing the desires ($Be^i$ and $Fe^i$) with the predictions ($B^i$ and $F^i$). It is not clear how $S^i$ and $Se^i$ differ. Once the designer is satisfied, she externalizes the functions, behaviors and structure ($F^i$, $B^i$ and $S^i$), creating the design specification ($F^e$, $B^e$, $S^e$), which enables manufacture of the design object.

Galle (2009) argued that the concepts of FBS may be interpreted in two ways. In a *nominalist interpretation*, $F$, $B$ and $S$ "stand for function-descriptions[,] ... behavior-descriptions, and structure-descriptions" (Galle 2009, p. 7) and FBS "becomes a model of a process of *symbol manipulation*" (Galle 2009, p. 7). In a realist interpretation, *F*, *B* and *S* are interpreted as "refer[ing] to *functions*, *behaviors* and *structures* that are regarded as *entities in their own right*" (Galle 2009, p. 8) – they exist in the same philosophical sense as numbers and constructs. Given the definition (above) of the external world as being composed of "representations", this paper adopts the nominalist interpretation.

FBS and Situated FBS are clearly process theories – they relate design to the lower-level activities shown in Table 3, have a (teleological) causal motor and Gero and Kannengiesser (2004) claimed that "the eight processes depicted in the FBS Framework are … fundamental for all designing" (p. 90). While FBS was intended originally to explain engineering design, Kruchten (2005) adapted FBS for software development and Gero and Kannengiesser (2007b) adapted Situated FBS to software development.

FBS is more consistent with the Rational Model in three ways. First, it includes neither construction nor deployment of the design object. Second, while it includes revision of the requirements ($FR^e$, $BR^e$ and $SR^e$) based on the realizations stemming from the design process, it does not include problem reframing based on changes to the context or other external forces. Third, it has an explicit focus on symbol manipulation.

The benefits of FBS include its meticulousness and the intuitive appeal of its main constructs – developers speak in terms of a software system's requirements (functions), behavior (features) and architecture (structure). Its limitations include the difficulty of explaining (especially Situated) FBS to non-experts and the absence of problem framing and deployment.

## 2.7 Summary and the Need for New Theory

Recalling the paper's motivation (§1), Brooks (2010) argued that education drives the need for a dominant, communicable model of software design and, despite its limitations, the Rational Model remains dominant because it is easy to understand and communicate. Therefore, breaking the Rational Model framing of the ISD discourse may spur new innovation in design thinking and overcome the limitations of the Rational Model. One way of doing this is to identify an alternative software design process theory consistent with the assumptions of the Empirical Model.

Consequently, this section reviews five of the most influential design process theories that appear at least partially applicable to software development. Four are consistent with the Rational Model while only Alexander's Selfconscious Process is consistent with the Empirical Model (Table 4). The need for new theory then is clear – while the Selfconscious Process may provide a starting point, it requires substantial clarification and elaboration to be relevant, useful and rigorously testable in the software development domain.

Table 4. Summary of Design Process Theories

| Theory | Proponent | Ideal Type | Perspective |
|---|---|---|---|
| The Waterfall Model | (Royce 1970) | Lifecycle | Rational |
| The Basic Design Cycle | (Roozenburg et al. 1995) | Lifecycle | Rational |
| Problem-Design Exploration Model | (Maher et al. 1995) | Evolutionary | Rational |
| Selfconscious Process | (Alexander 1964) | Teleological | Empirical |
| (Situated) FBS | (Gero 1990; Gero et al. 2004) (Gero and Kannengiesser 2007a; Kruchten 2005) | Teleological | Rational |

Before proposing new theory, however, it is crucial to understand what exactly the above theories claim. The hypotheses that constitute process theories may be less clear than the hypotheses that constitute variance theories as different types of process theories make different types of truth claims.

In Lifecycle theories including the Waterfall Model and the Basic Design Cycle, "the trajectory to the final end state is prefigured and ... [each stage] must occur in a prescribed order, because each piece sets the stage for the next. Each stage of development is seen as a necessary precursor of succeeding stages" (Van de Ven and Poole 1995, p. 515). Therefore, a lifecycle theory of process *P* posits a sequence of phases and claims not only that *P* can be divided into these phases but also that later phases are impossible without earlier phases. Contrastingly, "unlike life-cycle theory, teleology does not prescribe a necessary sequence of events or specify which trajectory development of the organizational entity will follow" (Van de Ven and Poole 1995, p. 516). Teleological theories including FBS and the Selfconscious Process posit an independent agent pursuing socially-constructed goals by engaging in activities in self-determined sequence. Therefore, a teleological theory of process P posits a goal-oriented, independent agent and a set of activities or activity categories from which the agent chooses.

As both theory types posit activities or phases, both may be incorrect in four ways, analogous to Wand and Weber's (2002) four types of modeling grammar deficiencies – phases or activities may be overloaded, redundant, extraneous or missing. In addition, a teleological theory may be deficient of it posits a non-existent agent and a lifecycle theory may be deficient if it posits an incorrect sequence (Table 5). For the purposes of this paper, one process theory is empirically superior to another if it exhibits fewer or less severe deficiencies. Of course, process theories have other quality dimensions including ease of applicability and level of abstraction. However, as developing a complete framework for evaluating process theories is beyond the scope of this paper, I focus on the above deficiencies.

Table 5. Possible Deficiencies in Teleological and Lifecycle Process Theories[1]

| Deficiency | Definition | Hypothetical Observation (gymnast example) |
|---|---|---|
| Process/Phase overload | Several loosely-coupled activites map into one Process or Phase | learning a headstand involves two relatively unconnected activities – strengthening the neck and improving balance |
| Process/Phase redundancy | Several Process or Phases map into one activity | some gymnasts simultaneously learn headstands and handstands through the same basic practice |
| Process/Phase excess | A Process or Phase does not map into any activity | some gymnasts learn headstands and handstands without learning the transition |
| Process/Phase deficit | An Activity does not map to any Process or Phase | all gymnasts learn tumbling while or between learning headstands and handstands to avoid injury |
| No Agent[2] | No actors map into the hypothesized independent, goal-seeking agent | rather than setting out to learn headstands or handstands, some gymnasts learn them as a byproduct of mastering the rings |
| Incorrect Sequence[3] | The activity sequence does not conform to the hypothesized sequence of phases | some gymnasts learn handstands before headstands |

[1] *Teleological theories have processes while Lifecycle theories have phases*

[2] *No agent applies only to teleological theories*

[3] *Incorrect sequence applies only to lifecycle theories*

# 3 PROPOSING SENSEMAKING-COEVOLUTION-IMPLEMENTATION THEORY

Following the need for a software design process theory (above), I elaborated Alexander's (1964) Selfconscious Process with concepts from IS and other design literature to create the *Sensemaking-Coevolution-Implementation* Theory (SCI) (Figure 7, Table 6). SCI's two core claims are that a complex software system is developed by an independent, goal oriented agent and that this agent engages in three basic processes – Sensemaking, Coevolution and Implementation – in a self-directed sequence.

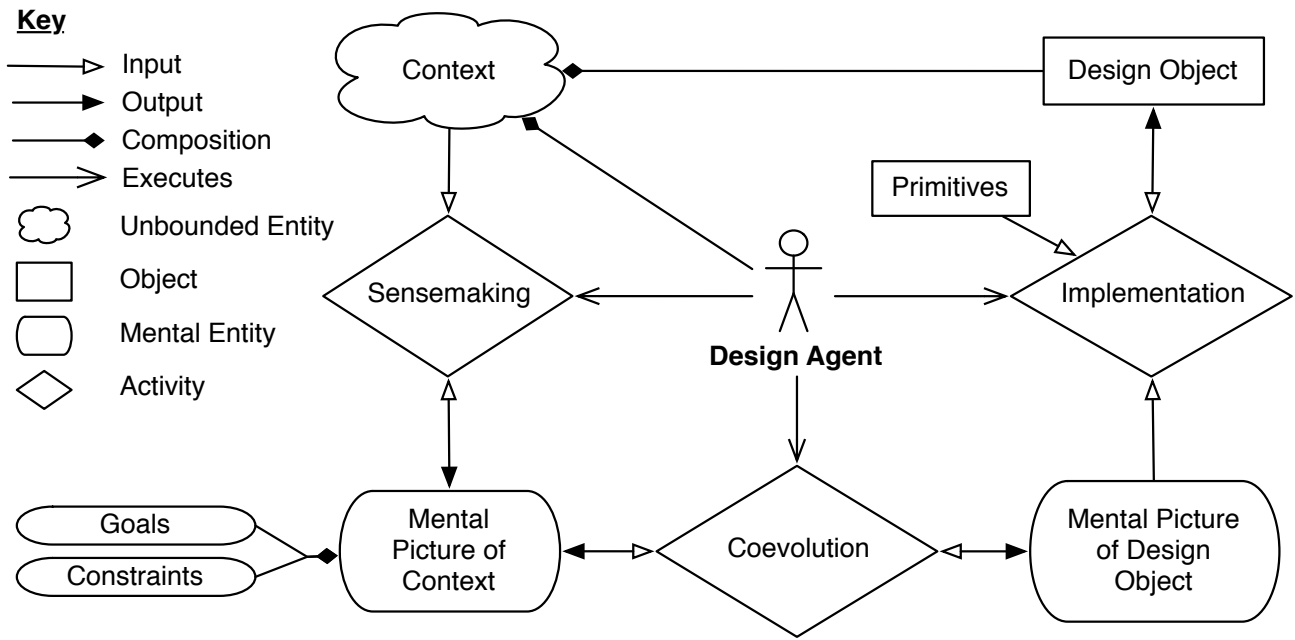

**Fig. 7.** Sensemaking-Coevolution-Implementation Theory

*Note: Arrows indicate relationships as shown, not a sequence of activities*

| Table 6. Concepts and Relationships of SCI, Defined ||
|---|---|
| Concept / Activity | Meaning |
| Constraints | the set of restrictions on the design object's properties |
| Design Agent | an entity or group of entities capable of forming intentions and goals and taking actions to achieve those goals and that specifies the structural properties of the design object |
| Context | the totality of the surroundings of the design object and agent, including the object's intended domain of deployment |
| Design Object | the thing being designed |
| Goals | optative statements about the effects the design object should have on its environment |
| Mental Picture of Context | the collection of all of the design agent's beliefs about its and the design object's environments |
| Mental Picture of Design Object | the collection of all of the design agent's beliefs about the design object |
| Primitives | the set of entities from which the design object may be composed |
| Sensemaking | the process where the design agent organizes and assigns meaning to its perception of the context, creating and refining the mental picture of context |
| Coevolution | the process where the design agent simultaneously refines its mental picture of the design object, based on its mental picture of context, and the inverse |
| Implementation | the process where the design agent generates or updates the design object using its mental picture of the design object |

## 3.1 A Detailed Illustration of SCI

To better communicate SCI, this section provides a detailed illustration. The following is not a case study or intended as evidence supporting SCI. It is merely an illustration of SCI using details from a real project.

PartnerPedia (www.partnerpedia.com) is an online partner management community where businesses find and build relationships with potential partner organizations (e.g., suppliers, distributors). In Summer 2008 the team had five members – AC and DA were professional web developers; JH was a computer science co-op student assisting with development; MG was the "product owner" and TB was a "quality assurance analyst". The team took a broadly Agile approach to development and loosely based its project management on Scrum (Schwaber 2004; Schwaber and Beedle 2001; Schwaber and Sutherland 2010). PartnerPedia was developed largely in-house as a new product rather than being designed for a particular client. The three core SCI processes were observable primarily through interpersonal interactions and modifications to the design object and boundary objects representing the mental pictures of the context and design object. Boundary objects including diagrams and prototypes are representational artifacts that are simultaneously robust enough to maintain their identities and flexible enough to serve multiple parties (Bergman et al. 2007; Star and Griesemer 1989).

The five team members collaborated closely, all working at the same time in the same room with short daily meetings to share progress and problems, i.e., acting together as a single agent. While the goal of the system was somewhat vague, the five team members shared the same basic idea that its purpose was to facilitate partner management, attract many users and be profitable by selling subscriptions for premium features. Upper management did not micromanage the team's process so the team could organize its activities independently. In summary, the five-person team constituted a goal-oriented, independent agent.

MG spent most of her time cajoling other firms into adopting, subscribing to or investing in PartnerPedia. As she spoke to dozens of (potential) users and stakeholders, she developed an increasingly detailed understanding of the diverse partnering and partner management processes used by different companies – one type of Sensemaking observed in the team. MG created boundary objects including notes and briefs that represented her mental picture of the context. As her understanding of the context increased, she generated more ideas about the product (her mental picture of the design object). She externalized these ideas in another boundary object, the "product backlog ... a prioritized list of everything that might be needed in the product". (Schwaber and Sutherland 2010, p. 5)

While this sounds like analysis followed by design (as in Waterfall) it was neither linear nor orderly. During planning meetings, MG shared the backlog items with the team, which discussed the high-level design of each item to estimate its effort and determine who would do it. To understand each item, team members would query MG about the context (another type of Sensemaking). The team's development of their understanding of the design object often triggered reconceptualization of the context. For example, different companies expressed the need for a company profile, a partner channel and a partner program, each of which had different meanings in the industry lexicon. However, as they devised and constructed each of these features, they realized that profiles, programs and channels were functionally equivalent. This realization triggered further reframing of the context and changes to their understanding of the product. All team members edited existing backlog items and wrote new ones as the team's collective understanding of the context evolved. Such coevolution of the team's understanding of the context and design object was most evident during planning meetings.

The team developed PartnerPedia in two-week *sprints*. AC, DA and JH spent most of these sprints expanding and refining the PartnerPedia website (Implementation). AC's work, for instance, was characterized by long periods of coding (Implementation) punctuated by breakdowns where AC would discover a non-trivial problem (e.g., a database design flaw creating errors wherever a company's name was displayed). Sometimes AC resolved breakdowns through Sensemaking activities including consulting programming documentation and examples. Other times breakdowns triggered a conversation with DA or MG. These conversations often involved sketching (cf., Prats et al. 2009) on a whiteboard and resulted in changes to boundary objects representing mental pictures of both the design artifact (i..e, the PartnerPedia site) and the context (e.g., the product backlog), indicating another kind of Coevolution.

The team relied on two kinds of testing – automated unit testing (cf., Beck 2002) (blackbox) and a quality assurance review at the end of each sprint (whitebox). Unit testing included determining what to test (Sensemaking) and coding the tests (Implementation). The team regularly ran the codebase against the test suite to check for failures. Unexpected failures triggered a search for errors, some of which resided in the code base while others resided in the test suite. Unit testing, therefore, involved Coevolution of code base and test suite. In contrast, near the end of each sprint TB would perform a quality assurance review on new and changed features. This whitebox testing effort consisted primarily of examining the design object and other aspects of the context (Sensemaking).

In summary, the PartnerPedia team functioned together as an independent, goal-directed agent. Behaviors associated with each of Sensemaking, Coevolution and Implementation were observable on both team and individual levels. At the team level, the SCI processes sometimes occurred in parallel while, at the individual level, team members alternated between them as necessary, sometimes switching processes several times in a few minutes. Again, PartnerPedia is used as an elaborate example to clarify SCI – it is not intended to support or justify SCI propositions.

## 3.2 Theoretical Justification of SCI

This subsection justifies both overall decisions regarding the structure of SCI and the specific concepts and relationships included. Section 3.4 discusses omitted concepts and relationships.

SCI was intentionally devised as a *teleological* process theory. This is consistent with characterizations of design by Churchman – "design belongs to the category of behavior called teleological, i.e., 'goal seeking' behavior" (1971, p. 5) and Löwgren et al. – "Our basic assumption about the design process is that its form, structure, and qualities are not given or ruled by laws of nature ... [but] by designers' own thoughts, considerations, and actions" (2004, p. 8). A teleological process theory is an explanation of how and why an entity changes where change is manifested by a goal-seeking agent engaging in activities in a self-determined sequence (Van de Ven and Poole 1995). Here, the changing entity is the design object (software system). 'Why' it changes is explained by adopting the teleological approach to causality (Gregor 2006; Kim 1999; Van de Ven and Poole 1995) – it changes because a person with free will chooses to change it. 'How' it changes is explained by the three SCI processes. Here, 'explain' is used in the reductionist sense – the higher-level activity (design) is explained by reducing it to a set of interconnected lower-level activities (Sensemaking, Coevolution, Implementation).

The first process is one where the design agent organizes his perceptions to create a meaningful mental picture of the context – "to convert a problematic situation to a problem, a practitioner must … make sense of an uncertain situation that initially makes no sense" (Schön 1983, p. 40). Schön's quotation raises suggests the construct *Sensemaking,* which is found in management (Weick 1995; Weick et al. 2005), knowledge management (Dervin 1998) and human-computer interaction (Furnas and Russell 2005; Russell et al. 1993). Sensemaking is simply "what people do to make sense of the information in their world" (Furnas and Russell 2005, p. 1). While no definition of Sensemaking has been widely accepted, it involves perceiving and encoding information to facilitate action (Russell et al. 1993) and noticing, bracketing, labeling, organizing and communicating ideas (Weick et al. 2005). For the purposes of SCI, Sensemaking generates, organizes and refines the mental picture of the context.

Sensemaking goes far beyond requirements elicitation as software development projects are affected by their entire social context (Orlikowski 1993a). Moreover, when a design agent engages in Sensemaking it may consider a variety of aspects of the project context, including aspects of the design agent's work environment, the design object's intended deployment domain and the environment in which the design object currently resides. This reflects the reality that internal factors including project budgets, interpersonal relationships between designers, managerial pressure all affect design decisions. The term "requirement" is omitted from SCI to avoid "the illusion of requirements" (cf. Ralph forthcoming).

The relationship between the mental pictures of context and design object is labeled *Coevolution* (after Maher et al. 1995). Coevolution refers to the mutual, iterative refinement of the design agent's mental pictures of context and design object. "In designing, 'the solution' does not arise directly from 'the problem'; the designer's attention oscillates, or commutes, between the two, and an understanding of both gradually develops" (Cross 1992, p. 5). A good example of this is provided by Dorst and Cross (2001):

> *A seed of coherent information was formed in the assignment information, and helped to crystallize a core solution idea. This core solution idea changed the designer's view of the problem. We then observed designers redefining the problem, and checking whether this fits in with earlier solution-ideas. Then they modified the fledgling-solution they had (p. 434).*

Many consider the idea of a designer mutually and iteratively refining mental pictures of the context (/ problem / environment / problem space) and the design object (/ form / solution space) central to design (e.g., Alexander 1964; Berente and Lyytinen 2006; Dorst and Cross 2001; Purao et al. 2002;

Schön 1983; Venable 2006). Important aspects of Coevolution include: 1) Schön's interconnected trio of problem framing, tentative adjustments to a design concept, and evaluating consequences; 2) organizing the system into components or subsystems (Alexander 1964; Parnas 1972) and 3) creativity (Dorst and Cross 2001; Wang and Ilhan 2009).

The relationship where the mental picture of the design object is realized in the actual design object is denoted 'Implementation' for consistency with the software industry vernacular (cf. Bourque and Dupuis 2004). It simply refers to creating and modifying the design object. Implementation may involve not only programming but also creating ancillary artifacts including installers, documentation and automated tests. Introducing the design object changes the context (Orlikowski 1993a), triggering further Sensemaking, Coevolution and Implementation.

Finally, SCI includes three concepts – goals, primitives and constraints – derived from Ralph and Wand's (2009) Design Process Input/Output Model. First, as teleological process theories posit a goal-oriented agent (Van de Ven and Poole 1995), SCI explicitly shows goals as part of the design agent's understanding of the context. Goals are also a central topic in the requirements engineering literature (Dardenne and Lamsweerde 1993). Second, primitives are the entities from which the design object is constructed (Ralph and Wand 2009). Primitives are included to recognize that software is not written like an essay; rather it is constructed from existing components, modules, libraries, design patterns, data structures, etc. Third, constraints are desired properties of and restrictions on the design object. Constraints are central constructs in both prescriptive and theoretical literature on software and engineering design (Boehm 1988; Bourque and Dupuis 2004; Gregor and Jones 2007; Lyytinen 1987; Sommerville 1996).

Like Alexander's Selfconscious Process (above), SCI is broadly consistent with the Empirical Model (Ralph 2011) in at least two ways. First, the Coevolution activity captures the design agent's oscillation between problem concept and solution concepts. Second, the explicit inclusion of context and Sensemaking reflects the view of the designer constructing rather than accepting goals and constraints. SCI also shares the Selfconscious Process' basic structure, with a context, a design object and mental pictures of each linked by three primary relationships. SCI expands on the Selfconscious Process, however, by clarifying the three core activities and including additional concepts crucial to software design. Each concept and relationship is grounded in existing research (Table 7).

**Table 7.** Theoretical Justification for SCI Concepts and Relationships

| Concept | Sources |
| --- | --- |
| Sensemaking | (Dervin 1998; Russell et al. 1993; Schön 1983; Weick 1995; Weick et al. 2005) |
| Coevolution | (Alexander 1964; Cross 1992; Dorst and Cross 2001; Maher et al. 1995; Schön 1983) |
| Implementation | (Alexander 1964; Boehm 1988; Bourque and Dupuis 2004; Royce 1970) |
| Constraints | (Ralph and Wand 2009; Simon 1996; Sommerville 1996) |
| Design Agent | (Alexander 1964; Cross 1992; Eekels 2000; Ralph and Wand 2009) |
| Context | (Alexander 1964; Checkland 1999; Orlikowski 1993a; Orlikowski 1993b; Ralph and Wand 2009; Schön 1983) |
| Design Object | (Alexander 1964; Eekels 2000; Ralph and Wand 2009) |
| Goals | (Churchman 1971; Dardenne and Lamsweerde 1993; Ralph and Wand 2009) |
| Mental Picture of Context | (Alexander 1964; Maher et al. 1995; Purao et al. 2002) |
| Mental Picture of Design Object | (Alexander 1964; Maher et al. 1995; Purao et al. 2002) |
| Primitives | (Meyer 1988; Parnas 1972; Ralph and Wand 2009; Robey et al. 2001) |

In summary, SCI explains how an agent designs a complex system by alternating between three categories of activities in a self-directed sequence. Its central claim is that Sensemaking, Coevolution and Implementation constitute a set of necessary and sufficient activities to generate a software system. Simply, an agent cannot create a software system without engaging in at least some Sensemaking, some Coevolution and some Implementation and there exists no activity strictly necessary for producing a software system that is not classifiable as Sensemaking, Coevolution or Implementation. Of course, a design agent may engage in numerous optional activities (below); however, the plethora of optional activities, concepts and artifacts a designer may use are omitted from SCI in the interests of parsimony.

## 3.3 Conceptual Evaluation

Based on the above discussion, several criteria for conceptually evaluating a new software design process theory are evident. Obviously, SCI can be evaluating using general criteria for theories including novelty, testability and usefulness. Furthermore, SCI was motivated by Cross's call for a simplifying paradigm for design and Brooks' identification of the need for a communicable model (§1), producing additional criteria: simplicity and communicability.

First, the Selfconscious Process exhibits one problematic limitation – its concepts and relationships are not clearly defined. Alexander notes that the interaction between the designer's conceptualizations of form and context (Coevolution) "contains both the probing in which the designer searches the problem for its major 'issues', and the development of forms which satisfy them; but its exact nature is unclear" (1964, p. 77). SCI is novel in its explicit definitions of these concepts and relationships. Moreover, while each concept and relationship previously existed in the literature, SCI combines them in a previously unseen way.

Second, SCI should produce testable propositions. In modern epistemology, testability is treated as a mutual property between two theories. Sober (1999) explains – "That some propositions are testable, while others are not, was a fundamental idea in the philosophical program known as logical empiricism. That program is now widely thought to be defunct" (p. 47); rather, "testing is an inherently contrastive activity – testing a hypothesis means testing it against some set of alternatives" (p. 52). Consequently, the testable propositions derivable from SCI depend on the process theory against which it is tested. SCI is certainly testable against FBS, for example, on at least the following dimensions:

- whether problem-setting and problem-solving are separate and sequential (FBS) or cotemporal and inextricably linked (SCI)
- whether the coding process is driven by prefigured decisions (FBS) or evolves iteratively with the design process (SCI)

- whether designers focus on models (FBS) or code (SCI)

Similarly, SCI is testable against the Waterfall Model on the first four dimensions of Table 5. For example, consider Waterfall's testing phase. Observing a test specialist applying several kinds of interconnected testing techniques in a manner loosely coupled with analysis, design or implementation activities would favor the Waterfall Model's view of testing and indicate activity overload or deficit in SCI. In contrast, suppose we observe two distinct kinds of testing – acceptance testing, where an analyst reviews a prototype with a client, and unit testing, where developers build test suites while building the software. Further suppose acceptance testing appears tightly coupled with other activities aimed at understanding the client's views and desiderata (Sensemaking), while unit testing appears tightly coupled with programming (Implementation). This would favor SCI's view of testing and suggest that Waterfall's "testing" phase is overloaded, i.e., it confounds two dissimilar phenomena.

In summary, SCI is empirically testable against other software development process theories in a variety of ways. Two ways to generate propositions are to examine the contrasting assumptions of the two theories and to examine specific deficiencies including process or phase redundancy, overload, excess or deficit. Although methodological advice for evaluating process theories is less abundant than for variance theories, innovation and change processes may be studied through questionnaires and field studies (Poole et al. 2000; Wolfe 1994). For questionnaires, items may be generated for each contrasting prediction on bipolar scales such that, for example, a 'strong disagree' would indicate beliefs consistent with SCI where 'strong agree' would indicate beliefs consistent with FBS. Meanwhile, collecting in-depth, longitudinal data from a small number of software development teams may provide a rich evidentiary base for each theory. This evidence may be compared using an a priori coding scheme based on the two theories to determine which is more veracious. Furthermore, combining the two approaches enables multi-method triangulation – the survey allows for random sampling and reliability while the field study facilitates gathering deep insights into developer behaviors and cognitive processes. While this paper primarily employs conceptual evaluation, due to the complexity of theoretically justifying SCI, this discussion demonstrates that SCI is testable in principle.

Third, if SCI is empirically veracious, it should be useful for research, education and practice. By providing a clear alternative, SCI clarifies the assumptions and implications of Rational Model process theories, facilitating analytical and empirical scrutiny. SCI may be empirically tested against FBS or the Waterfall Model to generate empirical insight into each theory's assumptions and deficiencies. From an education perspective, SCI provides an alternative to the Waterfall Model as a framework for teaching systems analysis and design. Furthermore, SCI can been used as a basis to evaluate courses and curricula. For instance, SCI has been used to analyze SE2004 and IS2010, the ACM and AIS model curricula for SE and IS undergraduate programs (Ralph 2012) could be analyzed for coverage of Sensemaking, Coevolution and Implementation and improved if Coevolution, for example, were poorly covered. From a practical perspective, SCI may facilitate evaluating and improving design methods, tools and practices. For example, in evaluating an SDM, we may question whether it provides guidance concerning all three fundamental design activities and, if not, can it be improved by considering those omitted?

Fourth, Cross (1992) elucidated the need for a "simplifying paradigm" for understanding and communicating design. Section 2 reviewed several design process theories, including the Selfconscious Process and the Situated FBS Framework. Imagining these theories on a scale ranging from too simple on the left too specific on the right, the Selfconscious Process appears perhaps too far left, i.e., so simple that it is difficult to draw useful propositions from it. In contrast, Situated FBS with its three worlds, twelve artifacts and twenty processes is perhaps too far right, i.e., so complex that it is difficult to understand. SCI is somewhere between, but still left of center. Favoring simplicity over specificity in SCI's development was an intentional choice, based on the recognition that, if successful, SCI will likely grow more complex as dimensions including power and time (§3.4) are added.

Fifth, Brooks (2010) argued that education drives the need for a communicable model of software design. SCI's communicability is enhanced by the clear definitions of each concept and relationship provided above. Anecdotally speaking, students and academics seem to understand my presentations of SCI quite readily, except for one common misunderstanding. Both students and academics continually mistake SCI for a method, as evidenced by questions including 'why would developers chose SCI over Waterfall?' and 'how do developers apply SCI?'. Of course, developers neither choose nor apply SCI anymore than technology adopters choose or apply the Technology Acceptance Model (cf., Venkatesh et al. 2003) as SCI is a theory, not a method.

### 3.4 Limitations of and Future Additions to SCI

The presentation of SCI given in this paper is limited in several important ways. First, the purpose of this paper is to present and theoretically justify SCI – empirical justification is left to future work. Any use of SCI therefore should be approached with caution as it has not yet been empirically evaluated.

Second, SCI focuses on activities strictly necessary for software design to occur. It does not comprehensively enumerate all design-related activities. For example, developers may externalize aspects of their mental pictures using boundary objects including notes and feature descriptions (§3.1). However, as writing notes is not strictly necessary for constructing software, SCI does not include a 'writing notes' activity. This is consistent with existing design process theories – none of the process theories reviewed above attempt to capture all possible optional design activities. Sim and Duffy (2003) proposed an "ontology of generic engineering design activities" including "abstracting", "decision making" and "searching". While mapping Sim and Duffy's ontology onto design process theories may be theoretically fruitful, attempting to include all 27 activities in the first version of a new process theory would likely make it confusing – a criticism previously lodged against Situated FBS (above).

Similarly, accounting for each aspect of design in one process theory would significantly increase its complexity. As all the other process theories reviewed above, SCI omits important concepts including environmental factors, individual differences, intra-agent dynamics, negotiation, politics, power, quality, success and time. Attempting to account for all these dimensions would hamper the theory's communicability and usefulness as a simplifying paradigm for understanding design, undermining the motivation given in the introduction. Therefore, exploring these and other dimensions of design is left to future work while the current theory focuses on explaining design behavior as simply and understandably as possible.

## 4 SCI AND ISD LITERATURE

SCI broadly contributes to design science, specifically the 'study of designers' research stream (cf. Hevner and Chatterjee 2010a; Ralph 2010b). This overlaps with the well-established research stream on information systems development.

Sambamurthy and Kirsch (2000) identified seven core concepts in ISD literature – *tasks* (of the development team including planning, analysis, design), *stakeholders*, *agenda* (stakeholders' goals), *transactions* (stakeholder actions), *context* (outside of the stakeholders and team), *structure* (e.g., SDMs, tools, policies) and *outcomes*. Similarly, Curtis (1988) proposed a "layered development model", which gives the possible units of analysis for studying ISD, including "individual", "team", "project", "company" and "business milieu". Meanwhile, Bansler (1989) organized (Scandinavian) ISD research into three traditions: the *systems theoretical*, which seeks to increase efficiency through technology; the *socio-technical*, which seeks to improve productivity and wellbeing through greater consideration of human factors; and the *critical*, which seeks improve the employee-employer power balance. From these perspectives, SCI explains how individuals and teams behave at the task level, while abstracting the remaining concepts and layers into the design context concept, from a viewpoint similar to Bansler's socio-technical perspective.

More generally, it is important to situate SCI in existing ISD literature. Exploring SCI's relationship to a selection of themes in ISD literature may better illuminate its nature and contribution. The remainder of this section discusses SCI's relationship to three themes in ISD – interpersonal interactions (communication, learning, coordination and negotiation), the trend toward distributed (including open source) development and ISD methodology. These themes were chosen simply because they are useful in this context – a comprehensive review of ISD literature is beyond the scope of this paper.

### 4.1 Interpersonal Interactions in ISD

Developing large systems involves communication, learning and negotiation processes (Curtis et al. 1988). Furthermore, as the number of stakeholders and participants increases, coordination becomes increasingly challenging. This raises questions concerning what we know about communication, learning, negotiation and communication and how they relate to SCI.

While knowledge is often defined as true, justified belief, Orlikowski (2002) found that "knowing is not a static embedded capability or stable disposition of actors, but rather an ongoing social accomplishment, constituted and reconstituted as actors engage the world in practice" (p. 250). Furthermore, "organizations fail to learn from their experience in systems development because of limits of organizational intelligence, disincentives for learning, organizational designs and educational barriers" (Lyytinen and Robey 1999, p. 85) and, more specifically, "when team members are immersed in a design activity, they are often unable (or unwilling) to acquire knowledge that cannot be immediately put to use" (Walz et al. 1993, p, 70). In summary, we know that acquiring knowledge in ISD projects is often difficult.

In ISD projects, knowledge is often exchanged in a dialectic process (Curtis et al. 1988; Walz et al. 1993), through discussions where criticism of old ideas generates new ideas and discussants revise beliefs. Many of these knowledge-producing interactions involve current or intended users of the design object. A climate that encourages user participation does not guarantee success but appears to help (Butler and Fitzgerald 1997). More specifically, user participation leads to critical encounters (Newman and Robey 1992; Robey and Newman 1996) and conflict/resolution cycles (Robey and Farrow 1982; Robey and Newman 1996; Robey et al. 1993) that may improve designs and change project trajectories. However, while some longterm ISD trends (e.g., the Agile movement) encourage increased user participation, others undermine it. For example, the trend toward COTS (instead of custom development) transfers power and participation away from users (Bansler and Havn 1994). Moreover, users may be expected to take responsibility for projects even when their involvement is shallow (Beath and Orlikowski 1994).

Coordination, communication and control are tightly interconnected concerns within ISD projects. Brooks (1995) argued that adding human resources to a software project may not increase its velocity partly due to increased coordination overhead for larger teams. Furthermore, as team size increases, informal, ad hoc coordination mechanisms become insufficient (Mockus et al. 2002). This insight is at the heart of document-based development methods, which seek to protect the organization from knowledge loss and miscommunication by recording crucial details in requirements specifications, software documentation and other boundary objects (e.g., Jacobson et al. 1999; Office of Government Commerce 2009). Indeed, team coordination is improved by shared mental models (Espinosa et al. 2001); however, Curtis et al. (1992) found that such documentation did not reduce the amount of communication required. Instead, "Artificial (often political) barriers to communication among project teams created a need for individuals to span team boundaries and to create informal communication networks" (Curtis et al. 1992, p. 1282). This partly explains designers' need for both vertical coordination (coordination between developers and users, through authorized entities including project managers) and horizontal coordination (between developers and users directly) (Nidumolu 1995). Geographic distance negatively effects coordination; however, this relationship is mediated by shared knowledge (Espinosa et al. 2007) (see §4.2). Finally, coordination issues are complicated by users, clients and managers who use formal and informal modes of control to influence development (Kirsch 1997; Kirsch et al. 2002; Nidumolu and Subramani 2003). Formal controls include defining appropriate business processes (behavioral control) and performance targets (outcome control). Informal controls include individuals setting their own targets (self-control) and groups that share values, approaches and goals (clan control).

SCI focuses on design *activities* from the perspective of the design agent. The term 'knowledge' was not used in SCI due to difficulty defining it. Rather, SCI posits that design agents generate, organize and revise beliefs and ideas about the project's context and design object through Sensemaking and Coevolution. Of course, whether the design agent is an individual or a team, one would expect complex intra-agent knowledge generation and communication processes. However, SCI sits at an intermediate level of abstraction between the individual and the organization. As such, intra-agent phenomena including the cognitive process of individual designers and coordination within design teams are not represented. Drilling down to a lower level of abstraction is left to future research.

Similarly, design projects may involve management processes (including coordination) parallel to the design process. Management activities are omitted from SCI not only to control scope but also as they are not inherent to design in the same way as SCI's activities. For example, consider a "guerrilla project" where well-meaning employees develop an application "under the radar" of management, relying on clan control (cf., Kirsch et al. 2002). Rather than a relationship between design agent and context, which could be shown in SCI, coordination in this case exists primarily within the design agent, i.e., at a lower level of abstraction (Curtis et al. 1988) than SCI. This caveat acknowledged, interconnections between management process and core design activities represent another potentially beneficial area of future research.

Considering user participation raises the related consideration of when a user becomes part of the design agent. Beck (2005) recommends having a user onsite at all times to answer questions from the design team. Such a user could conceivably become intermingled with the design team, sharing much of the team's collective mental model of the context and contributing to design decisions. At this point, the user is part of the design agent. This is related to the research stream on user-led systems design (Franz and Robey 1984b; Lawrence and Low 1993). The reverse question, when does the design team split into multiple agents, is discussed next.

## 4.2 The Scope of the Design Agent

Software development is increasingly distributed and global (Herbsleb and Moitra 2001). Distributed work takes more time than collocated work (Herbsleb et al. 2001) as it requires more people (Herbsleb and Mockus 2003) and coordination (Brooks 2010). Consequently, how to mitigate the productivity impact of distributed work has become a key concern in methods literature (cf., Sutherland et al. 2008). One form of distributed development that has received particular attention is open source software development, also called free or free/libra open source development (FLOSSD).

FLOSSD differs from traditional software engineering in its transparency (Scacchi et al. 2006), lack of formal SDMs, PMFs, budgets, schedules or rule structures (Scacchi 2007; Scacchi et al. 2006), developer self-assignment of tasks (Crowston et al. 2007) and success measures (Crowston et al. 2004; Crowston et al. 2006). For example, Crowston et al. (2006) argue that as the DeLone and McLean model of IS success (DeLone and McLean 1992; DeLone and McLean 2003) is based on development and deployment process very different from that observed in FLOSSD, the latter requires different success indicators including number of developers, level of development activity, individual reputation and movement from alpha to stable release.

FLOSSD projects are organized into layers – a small group of core developers, a larger group of ad hoc developers who make minor changes and bug fixes and an even larger group who report problems (Mockus et al. 2002). Core developers create extensive but incomplete shared mental models (Scozzi et al. 2008). Their adherence to the FLOSSD ideology enhances communication and trust, amplifying team effectiveness (Stewart and Gosain 2006). Crowston et al. (2005) found that team effectiveness was dominated by team composition and organizational context rather than tools or other technical factors. Rather than traditional forms of vertical and horizontal coordination (above; Nidumolu 1995), FLOSSD projects are coordinated using "informalisms" – boundary objects describing or prescribing aspects of the project including discussion forum threads, project wikis and to-do lists – which replace formalisms including requirements specifications and system documentation (Scacchi 2007).

Furthermore, Fitzgerald (2006) argued that FLOSSD is evolving from FLOSSD 1.0, where a core person or group of developers haphazardly developed a product with erratic support under GPL or a similar license, to FLOSSD 2.0 where major players are choosing open-source strategies and providing better support under a variety of licensing schemes. This shift may lead to substantive changes in or bifurcation of FLOSSD norms and patterns.

FLOSSD and distributed development more generally raise questions about SCI's structure and agent concept. Curtis et al. (1988) proposed a "layered development model" that identifies the multiple units of analysis (individual, team, project, company, business milieu) involved in ISD research. SCI uses the design agent concept to cross the individual and team levels – a design agent may be either an individual or a group of individuals *if and only if the group acts as a single agent*. A team may be considered a single agent when it acts in a coordinated way to achieve a shared goal under shared mental models. Simply, SCI assumes high team cohesion (cf., Beal et al. 2003). Obviously team cohesion may not be perfect; however, the key is to distinguish between reasonably cohesive teams and situations encountered in some distributed projects where an artifact is simultaneously developed by two or more separate teams exhibiting competition, rivalry, contradictory goals and weak coordination ties. Moreover, while a FLOSSD project's core developer group may act as a single agent, it would be incredulous to extend the design agent concept to the outer layers. Therefore, SCI does not currently generalize to FLOSSD or distributed development; furthermore, exploring the implications of multi-agent design is a potentially fruitful avenue for future research.

### 4.3 ISD Prescriptions - Methods and their Limitations

Avison and Fitzgerald (1999; 2003) divided the history of SDMs into four eras. In the "pre-methodology era", developers trained in programming but not the "contexts of use" built software from a superficial understanding of users' needs and goals, leading to many failures. In the "early methodology era", development was divided into phases inspired by the Waterfall model. In the "methodology era" proper, practitioners began using the term "methodology" and many approaches emerged including structured programming, prototyping, object-oriented development and participative development. In the (current) "post-methodology era", many practitioners have rejected methodologies in general. For example, developers may omit elements of methods not out of ignorance but because they perceive them as irrelevant in their context (Fitzgerald 1997). Possible explanations for methodology rejection include perceived ineffectiveness or poor usability of specific methods and perceived ineffectiveness of great methodicalness *in principle*. These explanations have starkly different implications – the first implies a need for better methods; the second implies a need for something other than methods.

Many studies continue to defend methods and methodicalness in principle. For example, Herbsleb et al. (1997) found that higher process maturity (i.e., greater methodicalness) is associated with higher product quality, customer satisfaction, productivity and morale, justifying the research program on Capability Maturity Model Integration (CMMI) and software process improvement (Aaen et al. 2001) more generally. Later, Staples et al. (2007) found that organizations do not adopt

CMMI despite believing in its benefits as they think they are too small, CMMI is too costly, or they lack time. More recently, the Agile project management framework Scrum has been linked to higher productivity (Cardozo et al. 2010). Several (quasi-scientific) surveys report that projects using Agile approaches have higher success rates than projects using linear or ad hoc approaches (Ambler 2008; Ambler 2010; VersionOne 2011).

However, many studies also argue that academic literature is biased toward formal methodologies (Fitzgerald 1996) and methodicalness and ignores "the possibility that amethodical development might be the normal way in which the building of [software] systems actually occurs in reality" (Truex et al. 2000, p. 58). *Amethodical* systems development refers to "management and orchestration of systems development without a predefined sequence, control, rationality, or claims to universality" (Truex et al. 2000, p. 54). Although the case for amethodical development is complex, at least three dimensions are evident:

1. Many practitioners do not use methods as prescribed (Bansler and Bødker 1993; Mathiassen and Purao 2002; Parnas and Clements 1986).
2. Several field studies found that practitioners acted in a fundamentally amethodical manner characterized by improvisation, opportunism and either absent or impotent formal controls (Baskerville and Pries-Heje 2004; Baskerville et al. 1992; Walz et al. 1993; Zheng et al. 2011).
3. In many situations, the desiderata – goals, requirements, requests, wants, etc. – are unknown, ambiguous, disputed, inarticulable, changing, misunderstood or conflicting (Brooks 2010; Curtis et al. 1988; Gladden 1982; Herbsleb et al. 2005; Royce 1970; Shenhar et al. 2001; Truex et al. 1999; Walz et al. 1993; Zheng et al. 2011).

This debate raises a host of questions including, do more methodical development efforts outperform less methodical development efforts? If so, which methods are best? If not, what can increase the effectiveness of amethodical development efforts? Additionally, to what extent are these relationships context-dependent; does system complexity moderate the relationship between methodicalness and performance?

SCI is a process theory, not a method; therefore, it makes no directly-applicable prescriptions. However, SCI may be used to analyze some types of prescriptions. For example, SCI posits that envisioning the design object changes the designer's understanding of the context. If correct, this implies that practices such as "freezing requirements" prior to design concept generation are counterproductive. Regarding the methodical/amethodical debate, SCI's assumptions are broadly more consistent with the worldview of amethodical development than with formal methods (especially document-driven methods). For example, the former involves a reactive, improvising agent (like SCI) while the latter (e.g., Jacobson et al. 1999) often assume known goals or imply that analysis, design and programming are temporally and physically separable (unlike SCI).

# 5 CONCLUSION

This paper was motivated by the following logic: 1) the design literature is bifurcated into a dominant (Rationalist) paradigm and a challenging (Empiricist) paradigm; 2) an effective challenging paradigm facilitates recognizing and ameliorating the limitations of the challenging paradigm; 3) the Empirical Model's effectiveness was limited by the lack of a simplifying process theory to counterbalance Rationalist theories like FBS. Therefore, this paper attempts to formulate a process theory that explains how complex software systems are created by collocated software development teams in organizations, consistent with the Empirical Model of Design.

The resulting theory, SCI, posits three primary design activities – Sensemaking, Coevolution and Implementation – enacted by an independent design agent in a self-determined sequence. Each of SCI's concepts and relationships are theoretically justified by reference to existing research. Therefore, rather than proposing new constructs, the paper's contribution lies in the innovative synthesis and clarification of existing constructs.

Due to the complexity of expressing and theoretically justifying the new theory, only conceptual evaluation of SCI is attempted. Empirical evaluation is left to future research. One approach would be to empirically contrast SCI against one or more other process theories (§2). For example, SCI may be tested against FBS and Waterfall as they imply different predictions (§3.3). Such evaluation is amenable to both field studies and questionnaires (Poole et al. 2000; Wolfe 1994).

If SCI is veracious, it is useful for research, practice and teaching. For research, SCI may facilitate evaluating and improving design methods, tools and practices. For example, in evaluating an SDM, we may ask "does this methodology provide guidance concerning all three fundamental design activities – Sensemaking, Coevolution and Implementation?". If not, can the methodology be improved by considering those omitted? SCI may also inform development of an antecedent theory of design project success. For example, SCI suggests that success may depend on the proportion of the design space explored through Coevolution, which may be affected by cognitive biases (cf., Stacy and MacMillan 1995) including confirmation bias (Oswald and Grosjean 2004) and status quo bias (Jost et al. 2004).

For developers and project managers, SCI has several implications. First, practices intended to increase the rationality the design process, for instance by imposing separate 'analysis; and 'design' phases and documents, may appear irrational to designers as the practices contrast with the designer's contextualized practice (cf., Stolterman 1991). Consequently, developers may fake adherence to apparently sensible design methods (Parnas and Clements 1986). Additionally, the cotemporality of problem understanding and problem solving presents serious challenges in producing accurate, upfront estimates of budgets and schedules, undermining the logic of fixed-price and -schedule contracts (cf., Brooks 2010).

From a teaching perspective, SCI provides an alternative to the Waterfall model as a way to introduce software development fundamentals and facilitates a theoretical rather than method-oriented approach to discussing amethodical and Agile approaches. SCI may also be useful for evaluating and updating courses and curricula. For instance, if the IS undergraduate model curriculum (Topi et al. 2010) lacked treatment of one of the fundamental design activities (likely Coevolution) this would indicate a potential avenue for improvement.

In conclusion, recent breakthroughs in design science have made possible the conceptual fusion and simplification represented by SCI. SCI provides a powerful alternative language for explaining software development practice and dispelling common misconceptions. It gives cohesion and form to the implications of Empiricist assumptions for ISD. However, SCI is only an initial attempt at formulating such a theory and is intended to spur productive debate on theoretical lenses for understanding software development practice.